\documentclass[12pt]{article}

\usepackage{axodraw}
\usepackage{graphicx}
\usepackage{feynarts}

\makeatletter

\def\paragraph{\@startsection{paragraph}{4}{\z@}{+2.00ex plus
 +1ex minus +.2ex}{1.5ex plus .2ex}{\it\normalsize}}

\def\section{\@startsection {section}{1}{\z@}{+3.0ex plus +1ex minus
  +.2ex}{2.3ex plus .2ex}{\normalsize\bf\boldmath}}
\def\subsection{\@startsection{subsection}{2}{\z@}{+2.5ex plus +1ex
minus +.2ex}{1.5ex plus .2ex}{\normalsize\bf\boldmath}}
\def\subsubsection{\@startsection{subsubsection}{3}{\z@}{+3.25ex plus
 +1ex minus +.2ex}{1.5ex plus .2ex}{\normalsize\it}}

\expandafter\ifx\csname mathrm\endcsname\relax\def\mathrm#1{{\rm #1}}\fi

\newcounter{saveeqn}

\@addtoreset{equation}{section}

\newcount\@tempcntc
\def\@citex[#1]#2{\if@filesw\immediate\write\@auxout{\string\citation{#2}}\fi
  \@tempcnta\z@\@tempcntb\m@ne\def\@citea{}\@cite{\@for\@citeb:=#2\do
    {\@ifundefined
       {b@\@citeb}{\@citeo\@tempcntb\m@ne\@citea
        \def\@citea{,\penalty\@m\ }{\bf ?}\@warning
       {Citation `\@citeb' on page \thepage \space undefined}}%
    {\setbox\z@\hbox{\global\@tempcntc0\csname
b@\@citeb\endcsname\relax}%
     \ifnum\@tempcntc=\z@ \@citeo\@tempcntb\m@ne
       \@citea\def\@citea{,\penalty\@m}
       \hbox{\csname b@\@citeb\endcsname}%
     \else
      \advance\@tempcntb\@ne
      \ifnum\@tempcntb=\@tempcntc
      \else\advance\@tempcntb\m@ne\@citeo
      \@tempcnta\@tempcntc\@tempcntb\@tempcntc\fi\fi}}\@citeo}{#1}}

\def\@citeo{\ifnum\@tempcnta>\@tempcntb\else\@citea
  \def\@citea{,\penalty\@m}%
  \ifnum\@tempcnta=\@tempcntb\the\@tempcnta\else
   {\advance\@tempcnta\@ne\ifnum\@tempcnta=\@tempcntb \else
\def\@citea{--}\fi
    \advance\@tempcnta\m@ne\the\@tempcnta\@citea\the\@tempcntb}\fi\fi}

\newcommand{\lsim}
{\mathrel{\raisebox{-.3em}{$\stackrel{\displaystyle <}{\sim}$}}}

\def\asymp#1%
{\mathrel{\raisebox{-.4em}{$\widetilde{\scriptstyle #1}$}}}

\def\Nequal#1%
{\mathrel{\raisebox{-.5em}{$\stackrel{=}{\scriptstyle\rm#1}$}}}
\newcommand{\dsl}[1]{\not \hspace{-0.7mm}#1}
\def\dsl{\mathpalette\make@slash}
\def\make@slash#1#2{\setbox\z@\hbox{$#1#2$}%
  \hbox to 0pt{\hss$#1/$\hss\kern-\wd0}\box0}

\def\beq{\begin{equation}}
\def\eeq{\end{equation}}
\def\beqar{\begin{eqnarray}}
\def\eeqar{\end{eqnarray}}
\def\barr#1{\begin{array}{#1}}
\def\earr{\end{array}}
\def\bfi{\begin{figure}}
\def\efi{\end{figure}}
\def\btab{\begin{table}}
\def\etab{\end{table}}
\def\bce{\begin{center}}
\def\ece{\end{center}}
\def\nn{\nonumber}

\def\text{\textstyle}

\def\al{\alpha}

\def\de{\delta}

\def\si{\sigma}

\def\reffi#1{\mbox{Figure~\ref{#1}}}

\def\refta#1{\mbox{Table~\ref{#1}}}

\def\citere#1{\mbox{Ref.~\cite{#1}}}
\def\citeres#1{\mbox{Refs.~\cite{#1}}}

\newcommand{\TeV}{\unskip\,\mathrm{TeV}}
\newcommand{\GeV}{\unskip\,\mathrm{GeV}}
\newcommand{\MeV}{\unskip\,\mathrm{MeV}}

\newcommand{\ri}{{\mathrm{i}}}

\newcommand{\Oa}{\mathswitch{{\cal{O}}(\alpha)}}

\def\mathswitchr#1{\relax\ifmmode{\mathrm{#1}}\else$\mathrm{#1}$\fi}

\newcommand{\PW}{\mathswitchr W}
\newcommand{\Pw}{\mathswitchr w}
\newcommand{\PZ}{\mathswitchr Z}

\newcommand{\PH}{\mathswitchr H}
\newcommand{\Pe}{\mathswitchr e}

\newcommand{\Pne}{\mathswitch \nu_{\mathrm{e}}}
\newcommand{\Pnebar}{\mathswitch \bar\nu_{\mathrm{e}}}
\newcommand{\Pd}{\mathswitchr d}

\newcommand{\Pu}{\mathswitchr u}

\newcommand{\Ps}{\mathswitchr s}

\newcommand{\Pc}{\mathswitchr c}

\newcommand{\Pb}{\mathswitchr b}

\newcommand{\Pt}{\mathswitchr t}

\newcommand{\Pep}{\mathswitchr {e^+}}
\newcommand{\Pem}{\mathswitchr {e^-}}

\def\mathswitch#1{\relax\ifmmode#1\else$#1$\fi}

\newcommand{\MW}{\mathswitch {M_\PW}}

\newcommand{\MZ}{\mathswitch {M_\PZ}}
\newcommand{\MH}{\mathswitch {M_\PH}}
\newcommand{\Me}{\mathswitch {m_\Pe}}

\newcommand{\Md}{\mathswitch {m_\Pd}}
\newcommand{\Mu}{\mathswitch {m_\Pu}}
\newcommand{\Ms}{\mathswitch {m_\Ps}}
\newcommand{\Mc}{\mathswitch {m_\Pc}}
\newcommand{\Mb}{\mathswitch {m_\Pb}}
\newcommand{\Mt}{\mathswitch {m_\Pt}}

\newcommand{\GZ}{\Gamma_{\PZ}}

\newcommand{\sw}{\mathswitch {s_\Pw}}

\newcommand{\GF}{\mathswitch {G_\mu}}

\def\solid{\raise.9mm\hbox{\protect\rule{1.1cm}{.2mm}}}
\def\dash{\raise.9mm\hbox{\protect\rule{2mm}{.2mm}}\hspace*{1mm}}

\def\ie{i.e.\ }

\newcommand{\eennh}{\Pep\Pem\to\nu\bar\nu\PH}
\newcommand{\eeneneh}{\Pep\Pem\to\Pne\Pnebar\PH}
\newcommand{\eenmnmh}{\Pep\Pem\to\nu_\mu\bar\nu_\mu\PH}

\def\draftdate{\relax}
\def\mda{\relax}
\def\mua{\relax}
\def\mla{\relax}
\def\draft{
\def\thtystars{******************************}
\def\sixtystars{\thtystars\thtystars}
\typeout{}
\typeout{\sixtystars**}
\typeout{* Draft mode!
         For final version remove \protect\draft\space in source file *}
\typeout{\sixtystars**}
\typeout{}
\def\draftdate{\today}
\def\mua{\marginpar[\boldmath\hfil$\uparrow$]%
                   {\boldmath$\uparrow$\hfil}%
                    \typeout{marginpar: $\uparrow$}\ignorespaces}
\def\mda{\marginpar[\boldmath\hfil$\downarrow$]%
                   {\boldmath$\downarrow$\hfil}%
                    \typeout{marginpar: $\downarrow$}\ignorespaces}
\def\mla{\marginpar[\boldmath\hfil$\rightarrow$]%
                   {\boldmath$\leftarrow $\hfil}%
                    \typeout{marginpar: $\leftrightarrow$}\ignorespaces}
\def\Mua{\marginpar[\boldmath\hfil$\Uparrow$]%
                   {\boldmath$\Uparrow$\hfil}%
                    \typeout{marginpar: $\uparrow$}\ignorespaces}
\def\Mda{\marginpar[\boldmath\hfil$\Downarrow$]%
                   {\boldmath$\Downarrow$\hfil}%
                    \typeout{marginpar: $\downarrow$}\ignorespaces}
\def\Mla{\marginpar[\boldmath\hfil$\Rightarrow$]%
                   {\boldmath$\Leftarrow $\hfil}%
                    \typeout{marginpar: $\leftrightarrow$}\ignorespaces}
\overfullrule 5pt
\oddsidemargin -15mm
\marginparwidth 29mm
}

\def\stars{\strut\leaders\hbox{*}\hfill\strut}
\def\starline{\hfil\strut\hfil\hbox to \textwidth {\stars}\hfil}

\begin{document}
\thispagestyle{empty}
\def\thefootnote{\fnsymbol{footnote}}
\setcounter{footnote}{1}
\null
\draftdate\hfill KA-TP-2-2003\\
\strut\hfill MPI-PhT/2003-03 \\
\strut\hfill PSI-PR-03-03\\
\strut\hfill LC-TH-2003-008\\
\strut\hfill hep-ph/0301189
\vfill
\begin{center}
{\Large \bf\boldmath
Electroweak radiative corrections to \\
single Higgs-boson production in $\Pep\Pem$ annihilation
\par} \vskip 2.5em
\vspace{1cm}

{\large
{\sc A.\ Denner$^1$, S.\ Dittmaier$^2$, M. Roth$^3$ and 
M.M.~Weber$^1$} } \\[1cm]
$^1$ {\it Paul-Scherrer-Institut, W\"urenlingen und Villigen\\
CH-5232 Villigen PSI, Switzerland} \\[0.5cm]
$^2$ {\it Max-Planck-Institut f\"ur Physik 
(Werner-Heisenberg-Institut) \\
D-80805 M\"unchen, Germany}
\\[0.5cm]
$^3$ {\it Institut f\"ur Theoretische Physik, Universit\"at Karlsruhe \\
D-76128 Karslruhe, Germany}
\par \vskip 1em
\end{center}\par
\vskip 2cm {\bf Abstract:} \par We have calculated the complete
electroweak ${\cal O}(\alpha)$ radiative corrections to the single
Higgs-boson production processes $\Pep\Pem\to \nu_l\bar\nu_l\PH$
($l=\Pe,\mu,\tau$) in the electroweak Standard Model.  Initial-state
radiation beyond $\Oa$ is included in the structure-function approach.
The calculation of the corrections is briefly described, and numerical
results are presented for the total cross section. In the $\GF$
scheme, the bulk of the corrections is due to initial-state radiation,
which affects the cross section at the level of $-7\%$ at high
energies and even more in the $\PZ\PH$ threshold region.  The
remaining bosonic and fermionic corrections are at the level of a few
per cent.  The confusing situation in the literature regarding
differing results for the fermionic corrections to this process is
clarified.
\par
\vskip 1cm
\noindent
January 2003
\null
\setcounter{page}{0}
\clearpage
\def\thefootnote{\arabic{footnote}}
\setcounter{footnote}{0}

\section{Introduction}
\label{se:intro}

The investigation of the mechanism of electroweak symmetry breaking in
general and of the Higgs boson in particular will be one of the main
tasks at future colliders. While the LHC will discover the Higgs
boson, if it exists and has no particularly exotic properties, its
complete profile can only be studied in the clean environment of an
electron--positron linear collider.  These studies require adequate
theoretical predictions including radiative corrections and
finite-width effects.

In $\Pep\Pem$ annihilation there are two main production mechanisms
for the Standard Model (SM) Higgs boson.  The cross section of the
Higgs-strahlung process, $\Pep\Pem\to\PZ\PH$, rises sharply at
threshold to a maximum a few tens of GeV above the threshold energy
$\MZ+\MH$ and then falls off as $s^{-1}$, where $\sqrt{s}$ is the
centre-of-mass (CM) energy of the $\Pep\Pem$ system.  In the W-boson
fusion process, $\Pep\Pem\to\Pne\Pnebar\PH$, the incoming $\Pep$ and
$\Pem$ each emit a virtual W~boson which fuse into a Higgs boson. The
corresponding cross section grows as $\ln s$ and thus is the dominant
production mechanism for large energies.

In lowest order, the Higgs-strahlung process has been studied in
\citere{Ellis:1975ap} and the vector-boson fusion process in
\citere{Jones:1979bq}. The $\Oa$ electroweak radiative corrections to
the process $\Pep\Pem\to\PZ\PH$ have been calculated by different
groups \cite{Fleischer:1982af} many years ago.  The electroweak
corrections to $\eennh$ have attracted a lot of interest recently.
In \citere{Kniehl:1995at} the leading corrections 
in the limit of a heavy top quark, as well as the corresponding
QCD corrections in ${\cal O}(\alpha_{\mathrm{s}}^n\GF\Mt^2)$
with $n=1,2$, have been worked out.
The contributions of fermion and sfermion loops in the Minimal
Supersymmetric Standard Model have been evaluated in
\citeres{Eberl:2002xd,Hahn:2002gm}; at first sight, however, the
results of the two calculations do not agree, not even on the
fermion-loop contributions in the SM.  A first calculation of the
complete $\Oa$ electroweak corrections to $\eennh$ in the SM has been
performed very recently \cite{Belanger:2002me}. Analytical results for
the one-loop corrections to this process have also been obtained by
another group \cite{Jegerlehner:2002es} as {\sl MAPLE} output, but a
numerical evaluation of these results is not yet available.

In this paper we present first results of a completely independent
calculation of the $\Oa$ electroweak corrections to the complete
process $\eennh$ in the SM. 
Details on this calculation will be given
elsewhere.  Here we sketch only the main ingredients.

\section{Method of calculation}
\label{se:calc}

We have calculated the complete $\Oa$ electroweak virtual and real
photonic corrections to the processes $\eeneneh$, $\nu_\mu\bar\nu_\mu
H$, and $\nu_\tau\bar\nu_\tau H$.  For $\eeneneh$, this includes both
the corrections to the Higgs-strahlung and the vector-boson fusion
processes, which are taken into account coherently.

The calculation of the one-loop diagrams has been performed in the
't~Hooft--Feynman gauge both in the conventional and in the
background-field formalism using the conventions of
\citeres{Denner:1993kt} and \cite{Denner:1994xt}, respectively.  The
renormalization is carried out in the on-shell renormalization scheme,
as described there.  The electron mass $\Me$ is neglected whenever
possible.

The calculation of the Feynman diagrams has been performed in two
completely independent ways, leading to two independent computer codes
for the numerical evaluation. Both calculations are based on the
methods described in \citere{Denner:1993kt}.  The tensor coefficients
of the one-loop integrals are recursively reduced to scalar
integrals with the Passarino--Veltman algorithm
\cite{Passarino:1979jh} at the numerical level.  The scalar integrals
are evaluated using the methods and results of
\citeres{Denner:1993kt,'tHooft:1979xw}, where ultraviolet divergences
are regulated dimensionally and IR divergences with an infinitesimal
photon mass.  The two calculations differ in the following points.  In
the first calculation, the Feynman graphs are generated with {\sl
  Feyn\-Arts} version 1.0 \cite{Kublbeck:1990xc}.  Using {\sl
  Mathematica} the amplitudes are expressed in terms of standard
matrix elements and coefficients of tensor integrals.  Tensor 5-point
functions have been evaluated both by applying the usual
Passarino--Veltman reduction and by using the direct reduction to
4-point integrals of \citere{Denner:2002ii}. While the results based
on the Passarino--Veltman algorithm become numerically unstable at the
phase-space boundary owing to the appearance of inverse Gram
determinants and could only be rescued by a careful extrapolation out
of the numerically safe inner phase-space domains, the direct
reduction of \citere{Denner:2002ii} avoids inverse leading Gram
determinants, rendering the results of this approach well behaved near
the phase-space boundary.  The whole calculation has been carried out
in the conventional and in the background-field formalism.  The second
calculation has been done with the help of {\sl Feyn\-Arts} version 3
\cite{Hahn:2000kx} and {\sl FormCalc} \cite{Hahn:1998yk}.  The
analytical expressions generated by {\sl FormCalc} were translated
into C code.  In order to eliminate 5-point tensor integrals the
interference of the pentagon diagrams with the lowest-order amplitude
was evaluated with {\sl FeynCalc} \cite{Mertig:an} by evaluating the
fermion traces. Then the loop momenta of the 5-point integrals
appeared in the numerator only in scalar products which could be
cancelled with denominators, leaving only scalar 5-point integrals.

The results of the two different codes, those obtained within the
conventional and background-field formalism, and those resulting from
different treatments of the tensor 5-point functions are all in good
numerical agreement (typically within at least 12 
digits for non-exceptional phase-space points).

\newcommand{\snn}{s_{\nu\bar\nu}}
We use two different schemes for the inclusion of the finite Z-boson
decay width. In the {\it fixed-width scheme}, each resonant Z-boson
propagator $1/(\snn-\MZ^2)$, where $\snn$ is the invariant mass of the
neutrino--antineutrino pair, is replaced by
$1/(\snn-\MZ^2+\ri\MZ\GZ)$, while non-resonant contributions are kept
untouched. As a second option, we applied a {\it factorization scheme}
where the full (gauge-invariant) $\PZ\PH$-production amplitude with
zero $\PZ$-boson width is rescaled by a factor
$(\snn-\MZ^2)/(\snn-\MZ^2+\ri\MZ\GZ)$. Within integration errors both
schemes give the same results for the total cross section.

The matrix elements for the real photonic corrections are evaluated
using the Weyl--van der Waerden spinor technique as formulated in
\citere{Dittmaier:1999nn} and have been successfully checked against
the result obtained with the package {\sl Madgraph}
\cite{Stelzer:1994ta}.  The soft and collinear singularities are
treated both in the dipole subtraction method following
\citeres{Dittmaier:2000mb,Roth:1999kk} and in the phase-space slicing
method following closely \citere{bo93}.  Beyond $\Oa$
initial-state-radiation (ISR) corrections are included at the
leading-logarithmic level using the structure functions given in
\citere{lep2repWcs} (for the original papers see references therein).

The phase-space integration is performed with Monte Carlo techniques
in both computer codes. The first code employs a multi-channel Monte
Carlo generator similar to the one implemented in {\sl RacoonWW}
\cite{Roth:1999kk,Denner:1999gp} and {\sl Lusifer}
\cite{Dittmaier:2002ap}, the second one uses the adaptive
multi-dimensional integration program {\sl VEGAS}
\cite{Lepage:1977sw}.

\section{Numerical results}
\label{se:numres}

\subsection{Input parameters}

For the numerical evaluation we use the following set of
SM parameters \cite{Hagiwara:pw},
\beq
\begin{array}[b]{lcllcllcl}
\GF & = & 1.16639 \times 10^{-5} \GeV^{-2}, \quad&
\alpha(0) &=& 1/137.03599976, \\
\MW & = & 80.423\GeV, &
\MZ^{\mathrm{LEP}} & = & 91.1876\GeV, &
\GZ^{\mathrm{LEP}} & = & 2.4952\GeV, \\
\Me & = & 0.510998902\MeV, &
m_\mu &=& 105.658357\MeV,\quad &
m_\tau &=& 1.77699\GeV, \\
\Mu & = & 66\MeV, &
\Mc & = & 1.2\GeV, &
\Mt & = & 174.3\;\GeV, \\
\Md & = & 66\MeV, &
\Ms & = & 150\MeV, &
\Mb & = & 4.3\GeV. 
\end{array}
\label{eq:SMpar}
\eeq
We do not calculate the W-boson mass from $\GF$ but use its
experimental value as input.  Since we employ a fixed width in the
resonant Z-boson propagator in contrast to the approach used at LEP to
fit the Z~resonance, where a running width is taken, we have to
convert the ``on-shell'' values of $\MZ^{\mathrm{LEP}}$ and
$\GZ^{\mathrm{LEP}}$, resulting from LEP, to the ``pole values''
denoted by $\MZ$ and $\GZ$ in this paper. The relation of the two sets
of values is given by \cite{Bardin:1988xt}
\beqar
\MZ &=& \MZ^{\mathrm{LEP}}/
\sqrt{1+(\GZ^{\mathrm{LEP}}/\MZ^{\mathrm{LEP}})^2} = 91.1535\GeV,
\nn\\
\GZ &=& \GZ^{\mathrm{LEP}}/
\sqrt{1+(\GZ^{\mathrm{LEP}}/\MZ^{\mathrm{LEP}})^2} = 2.4943\GeV,
\label{eq:zparam}
\eeqar
i.e.\ the difference is of formal two-loop order and numerically
hardly visible in the results presented below.  The masses of the
light quarks are adjusted to reproduce the hadronic contribution to
the photonic vacuum polarization of \citere{Jegerlehner:2001ca}. Since
we parametrize the lowest-order cross section with the Fermi constant
$\GF$ ($\GF$ scheme), \ie we derive the electromagnetic coupling
$\alpha$ according to $ \alpha_{\GF} = \sqrt{2}\GF\MW^2\sw^2/\pi$, the
results are practically independent of the masses of the light quarks.
Moreover, this procedure absorbs the corrections proportional to
$\Mt^2/\MW^2$ in the fermion--W-boson couplings and the running of
$\al(Q^2)$ from $Q^2=0$ to the electroweak scale.  In the relative
radiative corrections, we use, however, $\alpha(0)$ as coupling
parameter, which is the correct effective coupling for real photon
emission.

We always sum over all three neutrino species, \ie over the processes
$\eeneneh$, $\nu_\mu\bar\nu_\mu H$, and $\nu_\tau\bar\nu_\tau H$.
Besides the full cross section, denoted ``total'' in the plots, we
also give the cross section resulting from the $\PZ\PH$-production
channel and the $\PW\PW$-fusion channel separately, which are referred
to as ``ZH'' and ``WW'' contributions, respectively. In the
$\PZ\PH$-production channel we sum over the relevant contributions of
all $\nu\bar\nu\PH$ final states, which is equivalent to multiplying
the cross sections for $\eenmnmh$ by a factor 3.  This means that the
results shown for ``total'' and ``ZH+WW'' only differ by the
interference terms between the $\PZ\PH$ and $\PW\PW$ channels.  In the
results presented here, the ISR is convoluted only with the
lowest-order cross section.  We consider merely total cross sections
without any cuts; distributions will be discussed elsewhere.

For reference we give some numbers for the total cross section in
lowest order, $\si_\mathrm{tree}$, and including electroweak
corrections, $\si$, together with the relative corrections defined as
$\de=\sigma/\sigma_{\mathrm{tree}}-1$ in \refta{ta:xsection}.
\begin{table}
\bce
\begin{tabular}{cccc}
\hline
$\MH$ [GeV] & $\sigma_{\mathrm{tree}}$ [fb] &
$\sigma$ [fb] & $\de$ [\%] \\
\hline
115 &   92.64(2)~  &    85.01(8)~  &  $~{-8.2(1)}$        \\
150 &   68.17(2)~  &    61.76(5)~  &  $~{-9.4(1)}$        \\
200 &   41.800(9)  &    37.76(3)~  &  $~{-9.7(1)}$  \\ 
250 &   23.764(4)  &    20.97(1)~  &  ${-11.8(1)}$  \\ 
300 &   12.125(2)  &    10.478(6)  &  ${-13.6(1)}$  \\ 
350 &   5.2047(6)  &    ~4.264(2)  &  ${-18.1(1)}$  \\ 
\hline
\end{tabular}
\ece
\caption{Lowest-order cross section for $\eennh$ in the $\GF$-scheme,
  $\si_{\mathrm{tree}}$, cross section including full electroweak
  corrections, $\si$, and relative corrections $\de$ for various Higgs
  masses at $\sqrt{s}=500\GeV$}
\label{ta:xsection}
\end{table}
The last numbers in parentheses correspond to the Monte Carlo integration
error of the last given digit.

\begin{figure}
\centerline{\includegraphics[width=.5\textwidth]{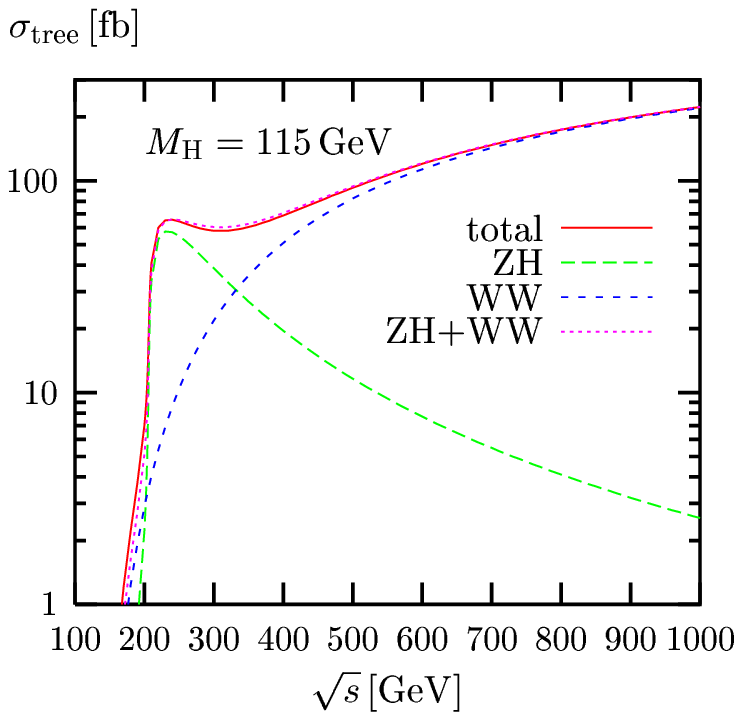}
\includegraphics[width=.5\textwidth]{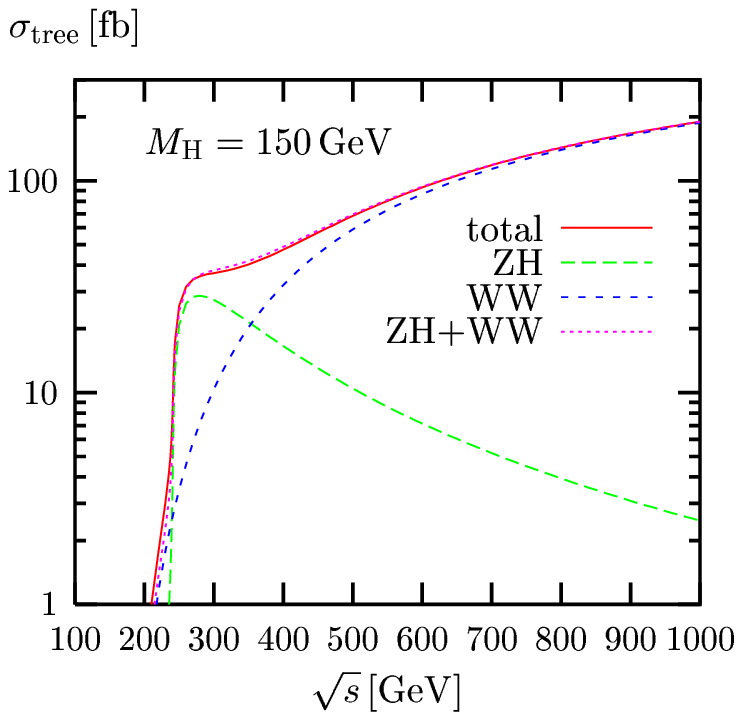}}
\caption{Lowest-order cross section and contributions from 
  $\PZ\PH$-production and $\PW\PW$-fusion channels for $\MH = 115
  \GeV$ and $\MH = 150 \GeV$}
\label{fi:born}
\end{figure}

In \reffi{fi:born} we show the lowest-order cross section as a
function of the CM energy for $\MH=115\GeV$ and $150\GeV$. Besides the
total contribution we separately give also the contributions from
$\PZ\PH$ production and $\PW\PW$ fusion as defined above.
\begin{figure}
\centerline{\includegraphics[width=.5\textwidth]{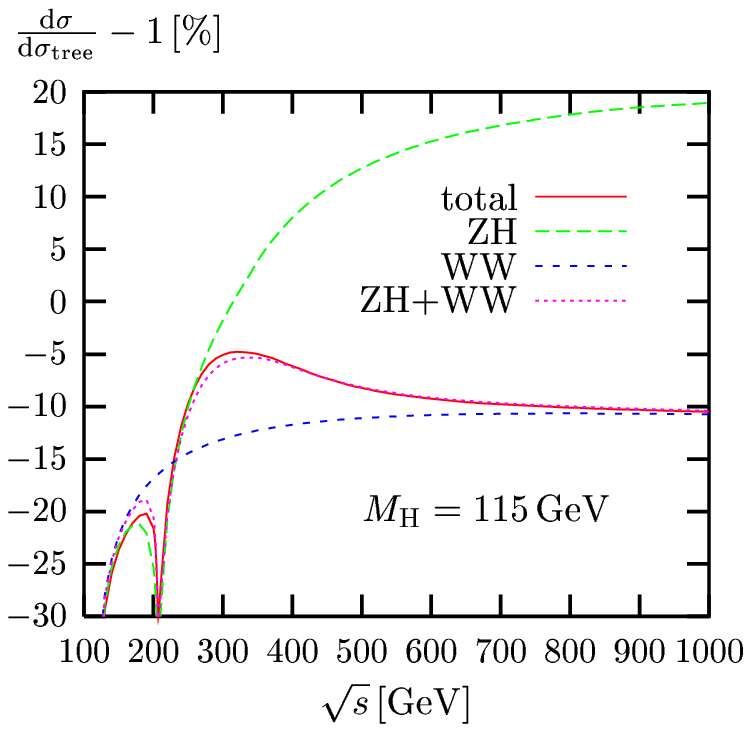}
\includegraphics[width=.5\textwidth]{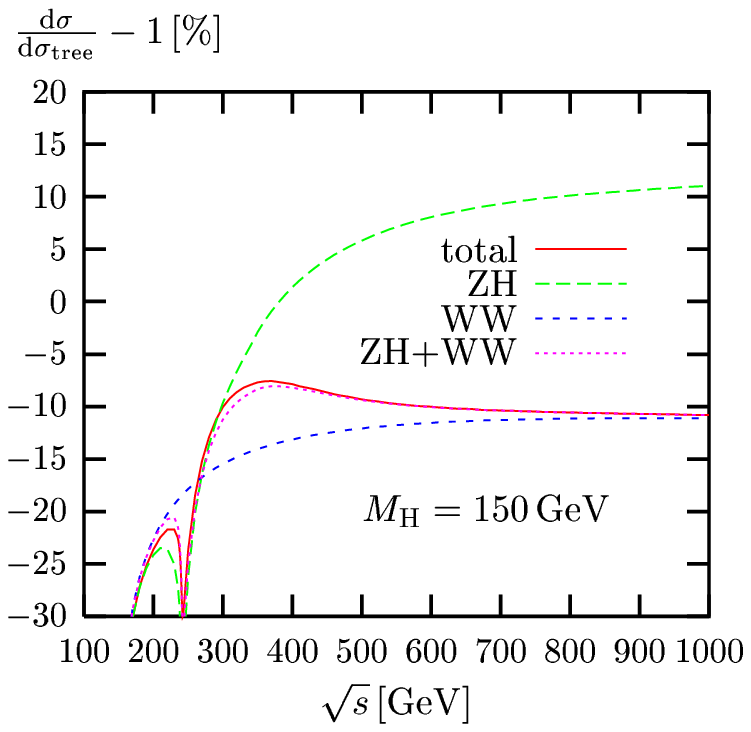}}
\caption{Relative electroweak corrections to the complete process
  $\eennh$ and to the contributions from $\PZ\PH$-production and
  $\PW\PW$-fusion channels for $\MH = 115 \GeV$ and $\MH = 150 \GeV$}
\label{fi:corr}
\end{figure}%
Below the $\PZ\PH$ threshold and above $400\GeV$ the cross section is
dominated by $\PW\PW$ fusion. The $\PZ\PH$-production contribution
dominates from the $\PZ\PH$ threshold up to about $300\GeV$.
Comparing the ``total'' cross section with the sum ``ZH+WW'', one can
see that the interference contributions between $\PZ\PH$ and $\PW\PW$
channels are small in lowest order for the inspected kinematical
situation.

The relative corrections to the lowest-order contributions of
\reffi{fi:born} are shown in \reffi{fi:corr}.  The corrections to the
$\PZ\PH$-production channel are large and negative ($\lsim -20\%$)
below threshold, rise fast above threshold, and reach 19\% and 11\% at
$\sqrt{s}=1\TeV$ for $\MH=115\GeV$ and $\MH = 150\GeV$, respectively.
The corrections to the $\PW\PW$-fusion channel are similar below the
$\PZ\PH$ threshold, rise sharply at the threshold and are flat and
about $-10\%$ above $500\GeV$.  The corrections to the complete
process are always negative. They follow those of the
$\PZ\PH$-production channel for energies below $\sim300\GeV$ and those
of the $\PW\PW$-fusion channel above $\sim500\GeV$.
 
\begin{figure}%
\centerline{\includegraphics[width=.5\textwidth]{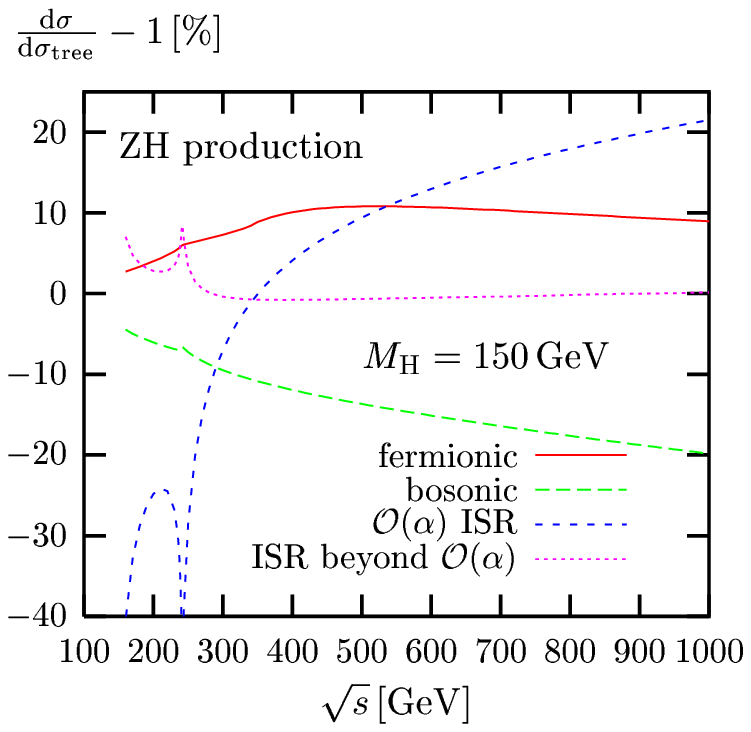}
\includegraphics[width=.5\textwidth]{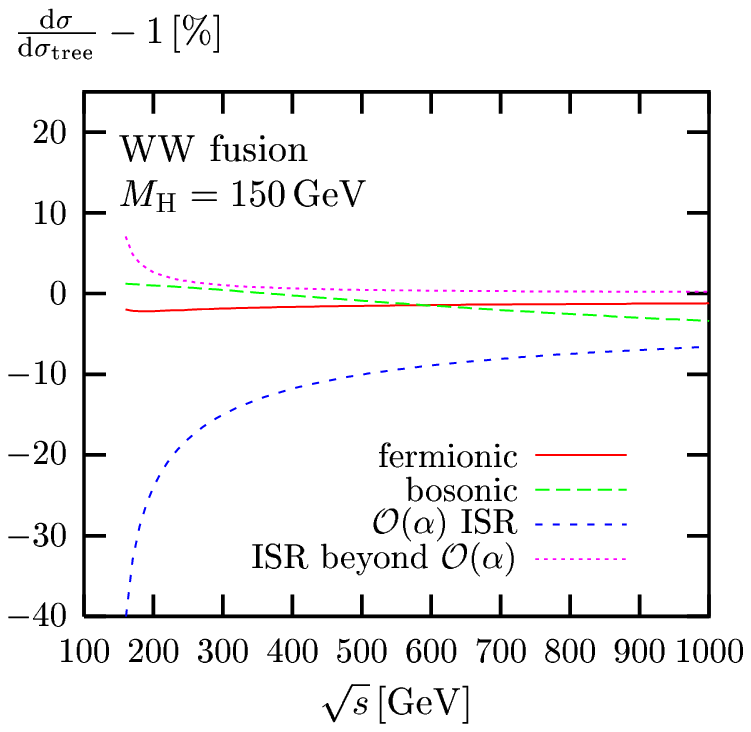}}
\caption{Relative electroweak corrections to the $\PZ\PH$-production
  (left) and $\PW\PW$-fusion (right) channels resulting from $\Oa$
  ISR, ISR beyond $\Oa$, fermion loops, and non-ISR bosonic
  corrections for $\MH = 150 \GeV$}
\label{fi:contrchannels}
\end{figure}

The contributions of $\Oa$ ISR corrections, ISR beyond $\Oa$,
fermionic corrections, and non-ISR bosonic corrections to the
$\PZ\PH$-production and $\PW\PW$-fusion channels are presented in
\reffi{fi:contrchannels}.
\begin{figure}
\centerline{
\includegraphics[width=.5\textwidth]{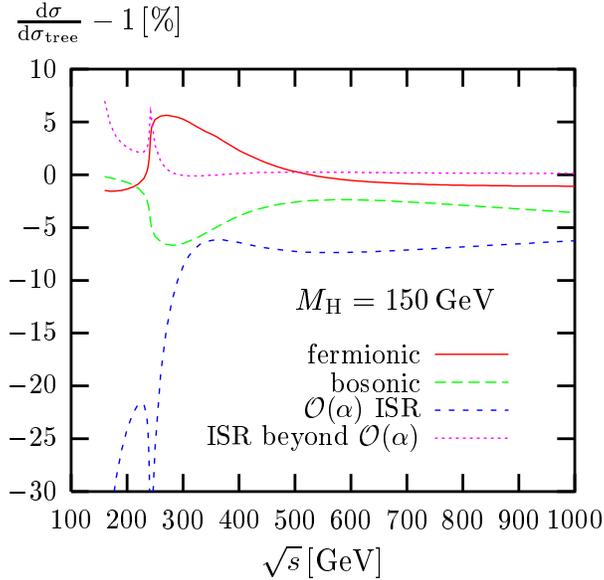}}
\caption{Relative electroweak corrections to the complete lowest-order cross
  sections resulting from $\Oa$ ISR, ISR beyond $\Oa$, fermion loops,
  and non-ISR bosonic corrections for 
  $\MH = 150 \GeV$}
\label{fi:contr}
\end{figure}%
The $\Oa$ ISR corrections are defined as the $\Oa$ contributions of
the structure functions of \citere{lep2repWcs}. The fermionic
corrections summarize all fermion-loop contributions to loop diagrams
and counter terms. The non-ISR bosonic corrections are obtained by
subtracting the $\Oa$ ISR corrections and the fermionic corrections
from the complete $\Oa$ corrections.  It can be seen, that the energy
dependence in both channels and, in particular, the rise of the
corrections to the $\PZ\PH$-production channel with energy is
essentially due to the ISR corrections. The large corrections above
the peak of the lowest-order $\PZ\PH$ cross section are due to the
decreasing cross section. Since the ISR corrections increase with the
height of this peak, they are larger for smaller Higgs masses (see
\reffi{fi:corr}).  The ISR corrections beyond $\Oa$ are at the level
of several per cent where the lowest-order cross section rises
strongly, but are below 1\% above $300\GeV$.  The non-ISR corrections
provide a measure of the genuinely weak corrections. For the
$\PZ\PH$-production channel (\reffi{fi:contrchannels} left) the
fermionic corrections increase slowly from 3\% to 11\% around
$500\GeV$ and then go down to $9\%$ at $1\TeV$. The non-ISR bosonic
corrections decrease from $-5\%$ to $-20\%$ over the considered energy
range.  This behaviour is typical if electroweak Sudakov logarithms of
the form $-\alpha\ln^2(s/\MZ^2)$ dominate the corrections, which is
always the case if the major part of the total cross section results
from intermediate scattering angles.  For the $\PW\PW$-fusion channel
(\reffi{fi:contrchannels} right), the ISR corrections are negative for
all energies, which can be explained by the rising cross section with
increasing scattering energy. ISR beyond $\Oa$ influences the cross
section only at the level of $\lsim1\%$ for energies above $300\GeV$.
The fermionic corrections are between $-1\%$ and $-2\%$ over the whole
energy range.  The non-ISR bosonic corrections fall from $+1\%$ at
threshold to $-3\%$ at $\sqrt{s}=1\TeV$. This energy dependence is
much weaker than for the $\PZ\PH$ channel, since the cross section for
the $\PW\PW$ channel is more and more dominated by small scattering
angles, i.e.\ not by the Sudakov regime.

The various contributions of the corrections to the complete process
are depicted in \reffi{fi:contr}.  The size of the corrections can be
easily explained by the results of \reffi{fi:contrchannels}; it is
given by the size of the corrections of the dominating channel.  The
ISR corrections vary strongly in the region of the $\PZ\PH$ threshold
but are nearly flat and about $-7\%$ for energies above $400\GeV$.
They are always negative since the lowest-order cross section is
continuously rising. The fermionic corrections reach a maximum of
about $6\%$
in the region where the $\PZ\PH$-production channel dominates and are
between $0\%$ and $-1\%$ above $500\GeV$ where the $\PW\PW$-fusion
process dominates. The non-ISR bosonic corrections exhibit a minimum
of about $-7\%$ where the $\PZ\PH$-production channel dominates and
are between $0\%$ and $-4\%$ elsewhere.

\section{Comparison with other calculations}
\label{se:comp}

We have compared our results for the $\Oa$ corrections to
\citere{Belanger:2002me} and the contributions from closed fermion
loops with \citeres{Eberl:2002xd,Hahn:2002gm}.

Adapting the input parameters and the parametrization of the
lowest-order matrix element to those used in \citere{Belanger:2002me},
we reproduced the numbers in Table 2 for the total cross section given
in the first paper of \citere{Belanger:2002me}. We find agreement
within 0.2\% for the total lowest-order cross section%
\footnote{As we were told by the F.~Boudjema, the integration errors
  of the numbers for the lowest-order cross section in Table~2 of
  \citere{Belanger:2002me}, which were suppressed in the table, are
  also of the order of $0.2\%$. When increasing statistics the
  agreement becomes better than $10^{-4}$ for the lowest-order cross
  section.}  and within 0.3\% for the corrected cross section. The
corrections relative to the lowest-order cross section agree within
0.2\%.  Note that the authors of \citere{Belanger:2002me} use
$\alpha(0)$ to parametrize the lowest-order cross section.  As a
consequence their relative corrections are shifted by $3\Delta
r\approx +9\%$ compared to those in the $\GF$ scheme.

Adapting the input parameters to those of \citere{Hahn:2002gm}, who
also use the $\GF$ scheme, we perfectly reproduce the SM results of
these authors for the tree-level cross section and the fermionic
corrections.

The authors of \citere{Eberl:2002xd} use $\hat\alpha(\MZ)$, \ie the
running electromagnetic coupling in the $\overline{\mathrm{MS}}$
scheme at $\MZ$, as defined in Eq.~(B.2) of \citere{Eberl:2001vb}, to
parametrize the lowest-order cross section.  The relative fermionic
corrections in this scheme differ from those in the $\GF$ scheme by
$3[(\Delta r)_{\mathrm{ferm}}-\Delta\hat\alpha(\MZ)]\approx -13\%$.
Moreover, Eberl et al.\ take into account only the loops from the
top--bottom doublet and omit the corrections to the
$\PZ\PH$-production channel. Adapting to this setup, we find agreement
with the results of \citere{Eberl:2002xd} at $1\TeV$ within the
integration errors for the lowest-order cross section.  The
corrections relative to the lowest-order cross section agree typically
within 0.3\%.  The large differences in relative corrections between
\citeres{Eberl:2002xd} and \cite{Hahn:2002gm} thus result essentially
from the use of different parametrizations of the lowest-order matrix
element.  The fact that the lowest-order SM cross sections in
\citeres{Eberl:2002xd} and \cite{Hahn:2002gm} agree qualitatively
despite of the different input-parameter schemes used is accidental
and due to the (different) input parameters.

Where the WW-fusion channel dominates, the non-ISR, \ie the genuine
weak, corrections are large if $\alpha(0)$ or $\alpha(\MZ)$ are used
to parametrize the lowest-order cross section. Only when using $\GF$
the large corrections associated with the running of $\al$ {\em and}
those proportional to $\Mt^2$ in the W-boson--fermion couplings are
absorbed in the lowest-order cross section.  For the ZH-production
channel the situation is more complicated and a dedicated improved
Born approximation is under investigation.  It should be noted that
higher-loop corrections become relevant in schemes where the one-loop
corrections are large. Since ISR corrections, the running of $\al$,
and the $\Mt$-dependent corrections are independent, one has to deal
with all these terms separately.

\section{Summary}
\label{se:sum}
We have presented results from a calculation of the complete
electroweak ${\cal O}(\alpha)$ radiative corrections to the single
Higgs-boson production process $\Pep\Pem\to \nu\bar\nu\PH$ in the
electroweak Standard Model. We find that the ISR corrections are of
the order of $-7\%$ at high energies and more than $-10\%$ near the
$\PZ\PH$ threshold. In this region even the higher-order ISR
corrections reach several per cent.  This is due to the strong energy
dependence of the lowest-order cross section. The non-ISR corrections
are at the level of a few per cent if the lowest order matrix element
is parametrized with the Fermi constant $\GF$. In other schemes these
corrections are of the order of 10\%. It has been pointed out that the
confusion in the literature regarding the size of the corrections to
$\eennh$ is due to the use of different schemes and input parameters.

\section*{Acknowledgement}

We thank H.~Eberl for providing us with the input parameters and the
precise definition of the charge renormalization used in
\citere{Eberl:2002xd} as well as some numbers for comparison.  T.~Hahn
is gratefully acknowledged for his effort in the comparison of the
results \cite{Hahn:2002gm} for the pure fermion-loop contributions
with ours.  Finally, we thank F.~Boudjema for providing us with the
complete input used in \citere{Belanger:2002me} and with improved
numbers for comparison.
This work was supported in part by the Swiss Bundesamt f\"ur Bildung und
Wissenschaft and by the European Union under contract
HPRN-CT-2000-00149.

\end{document}